# Morphological characterization of Ge ion implanted $SiO_2$ matrix using multifractal technique

R. P. Yadav[1*], V. Baranwal[2,], Sunil Kumar[3], Avinash C Pandey[2], A. K. Mittal[4]

[1]Department of Physics, Motilal Nehru National Institute of Technology, Allahabad 211 004, INDIA

[2]Nanotechnology Application Centre, University of Allahabad, Allahabad 211 002, INDIA

[3]Inter University Accelerator Centre, Aruna Asaf Ali Marg, New Delhi 110067, INDIA

[4]Department of Physics, University of Allahabad, Allahabad 211002, INDIA

[*]Email: aurampratap@gmail.com, ramyadav2006@rediffmail.com

## Abstract

200 nm thick $SiO_2$ layers grown on Si substrates and Ge ions of 150 keV energy were implanted into $SiO_2$ matrix with two fluences of $2.5 \times 10^{16}$ and $7.5 \times 10^{16}$ ions/cm$^2$. The implanted samples were annealed at 950$^o$ C for 30 minutes in Ar ambience. Topographical studies of implanted as well as annealed samples were captured by the atomic force microscopy (AFM). Two dimension (2D) multifractal detrended fluctuation analysis (MFDFA) based on the partition function approach has been used to study the surfaces of ion implanted and annealed samples. The partition function is used to calculate generalized Hurst exponent with the segment size. Moreover, it is seen that the generalized Hurst exponents vary nonlinearly with the moment, thereby exhibiting the multifractal nature. The multifractality of surface is pronounced after annealing for the surface implanted with fluence $7.5 \times 10^{16}$ ions/cm$^2$.

# 1. Introduction:

Semiconducting quantum dots (QDs) embedded in a thin film dielectric matrix have significant technological applications in various fields such as electronics, optoelectronics, and photonics. Due to their unusual characteristic QDs can be incorporated in semiconductor devices like light emitting diodes [1], lasers [2], and field effect transistors [3]. Ge nanocrystals embedded in dielectric materials have been widely considered for charge storage in nanocrystals gate memories in non-volatile memory devices [4].

In particular the ion implantation is a favourable technique, because of precise control over depth and concentration distribution of implanted ions [5, 6]. It also suggests massive flexibility in the QD formation by control of the process parameters like thermodynamical limitations, and extreme chemical purity for the synthesis of isotope pure Ge QDs [7].

Surface plays a significant role in various physical processes and different fields of modern day science and technology. It illustrates rough behaviour on nano-scale, so its characterization is very important [8]. For this purpose scaling law analysis is highly needed. The most common scaling is referred as self-affine scaling which is reminiscent of fractal. The progress of fractal and multifractal theory opens new ways of understanding image analysis [9]. Fractal is characterized by single scaling exponent while multifractal is generalization of such concept. The traditional multifractal analysis (MFA) is limited to stationary object measurements but cannot deal with those that are nonstationary. The detrended fluctuation analysis (DFA) has been developed to solve this inability which eliminates a certain trends hidden in the characterizing object [10]. The performance of the DFA was proved superior to the wavelet method when dealing with the nonstationary multifractal measure [11]. Both DFA and its multifractal version (MF-DFA) [12] have been used to resolve various nonstationary time series in many research fields such as nature of flame images [13], texture representation [14], tumor recognition in magnetic resonance images [15], and thin films images [8,16]. Multifractal analysis can help to understand the relationship between material structure and its properties. The aim of this work is to report the multifractal analysis of Ge quantum dots embedded into $SiO_2$ matrix by ion implantation method with varying fluences. Effect of annealing is discussed on multifractal properties of implanted Ge quantum dots into $SiO_2$ matrix.

## 2. Experimental methods:

Ge ions of 150 keV energy were implanted at room temperature into thermally grown $SiO_2$ layers with varying fluences from $2.5 \times 10^{16}$ to $7.5 \times 10^{16}$ ions/cm$^2$. Annealing of implanted samples were done at 950° C for 30 minutes in Ar atmosphere for defect removal induced by ion implantation. Surface morphology was studied with Nanoscope IIIA for atomic force microscopy (AFM) measurements in tapping mode and digitalized into $512 \times 512$ pixels.

## 3. Results and discussion:

Fig.1 (a, c) shows the surface morphological AFM images of a $SiO_2$ film on a Si substrate implanted with 150 keV, Ge ions at fluences of $2.5 \times 10^{16}$ before and after annealing and Fig. 1 (b, d) shows for the fluence of $7.5 \times 10^{16}$ ions/cm$^2$ before and after annealing. These images are of the size of $5\mu m \times 5\mu m$. The surface properties are affected with surface roughness at nano-meter scale; hence, its characterization is very necessary. Classically, it is characterized by average roughness ($R_a$) and interface width ($w$). $R_a$ is described as the estimated value of the surface relative to the plane of the surface and is given by $R_a = \langle |h(i,j) - \langle h(i,j) \rangle| \rangle$. $w$ is defined as standard deviation of height fluctuation and is defined as $w^2 = \langle \{h(i,j) - \langle h(i,j) \rangle\}^2 \rangle$. Here, $h(i,j)$ is the height fluctuation of surface at point $(i,j)$ measured by AFM and $\langle h(i,j) \rangle$ is the average over $N \times N$ points. The computed values of surface roughness ($R_a$ and $w$) of Ge ion implanted $SiO_2$ surfaces before and after annealing are listed in Table I and II.

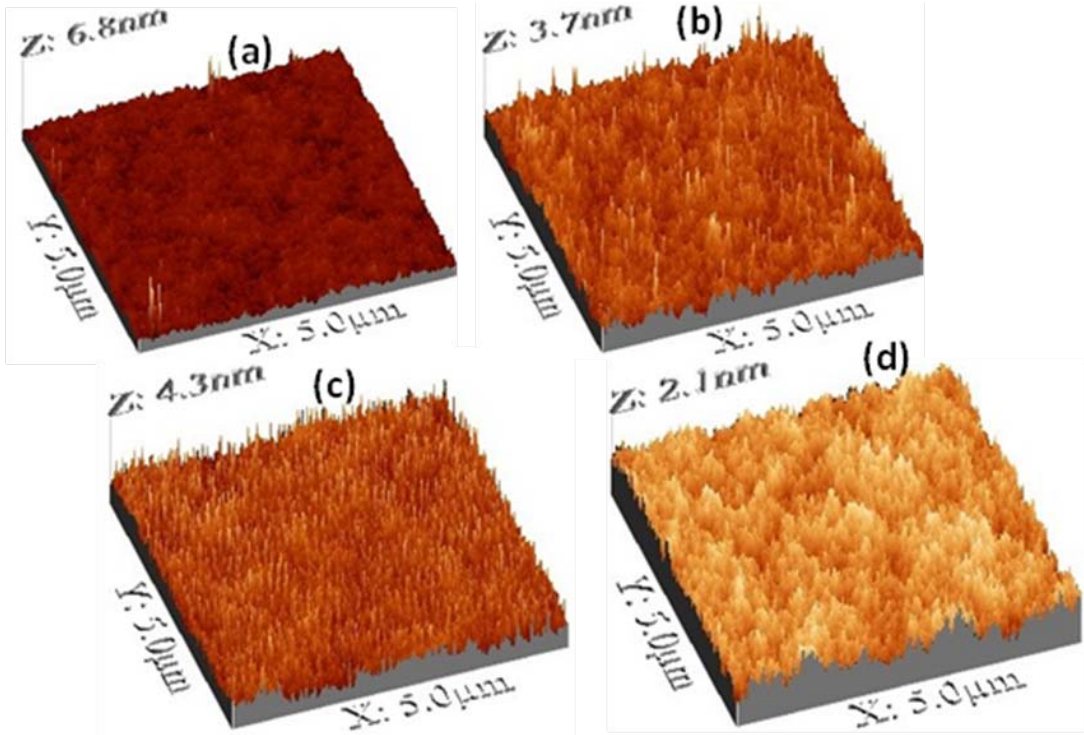

**Figure 1:** AFM images of embedded Ge nanocrystals in $SiO_2$ matrix implanted with fluences of $2.5 \times 10^{16}$ and $7.5 \times 10^{16}$ ions/cm$^2$ (a-b) and after annealing (c-d).

From Table I, it is clear that the roughness values are increased with implantation while decreasing nature is observed after annealing in Table II. These parameters are global characterizations of roughness and sensitive only to peak and valley values of surface fluctuations, missing the information about correlation and irregularity/ complexity about the surface.

**Table I:** Parametric values before annealing

| Fluence ions/cm$^2$ | $R_a$ [nm] | $w$ [nm] | $\alpha_{min}$ | $\alpha_{max}$ | $\Delta\alpha$ | $f(\alpha_{max})$ | $f(\alpha_{min})$ | $\Delta f$ |
|---|---|---|---|---|---|---|---|---|
| $2.5\times10^{16}$ | 0.1878 | 0.2450 | 1.9313 | 2.0861 | 0.1548 | 1.7121 | 1.7676 | 0.0166 |
| $7.5\times10^{16}$ | 0.2218 | 0.2947 | 1.9232 | 2.1002 | 0.1769 | 1.6788 | 1.7451 | 0.0277 |

**Table II:** Parametric values after annealing

| Fluence Ions/cm$^2$ | $R_a$ [nm] | $w$ [nm] | $\alpha_{min}$ | $\alpha_{max}$ | $\Delta\alpha$ | $f(\alpha_{max})$ | $f(\alpha_{min})$ | $\Delta f$ |
|---|---|---|---|---|---|---|---|---|
| $2.5\times10^{16}$ | 0.2966 | 0.4118 | 1.9480 | 2.0840 | 0.1360 | 1.7092 | 1.8492 | 0.1204 |
| $7.5\times10^{16}$ | 0.1968 | 0.2465 | 1.9398 | 2.1823 | 0.2425 | 1.3532 | 1.8047 | 0.4405 |

It is observed that surface height fluctuation often show self-affine properties [16]. A self-affine surface behaves like a fractal. Therefore, fractal analysis is needed for characterization of such surfaces.

The height-height correlation function or power spectral gives only the information about the fractal property of surface, which is a single exponent and is not enough to describe the dynamics of the surface [8]. To explore the multifractal nature of the Ge ion implanted SiO$_2$ surfaces before and after annealing, we applied the multifractal detrended fluctuation analysis (MFDFA). It is a powerful tool for detection of surface multifractlity [9]. The detail mathematical technique of MFDFA formalism is described in Ref. [8]. When MFDFA is used to investigate the multifractal scaling of object fluctuations, the method can assume partition function which can appear in double log plot of the fluctuation function $F_q(n)$ versus scale size $n$ for the different value of $q$ in the range $-7 < q < 7$. The slope of $F_q(n)$ versus $n$ is estimated using the principle of least square fit and is known as scaling exponent as shown in Fig. 2.

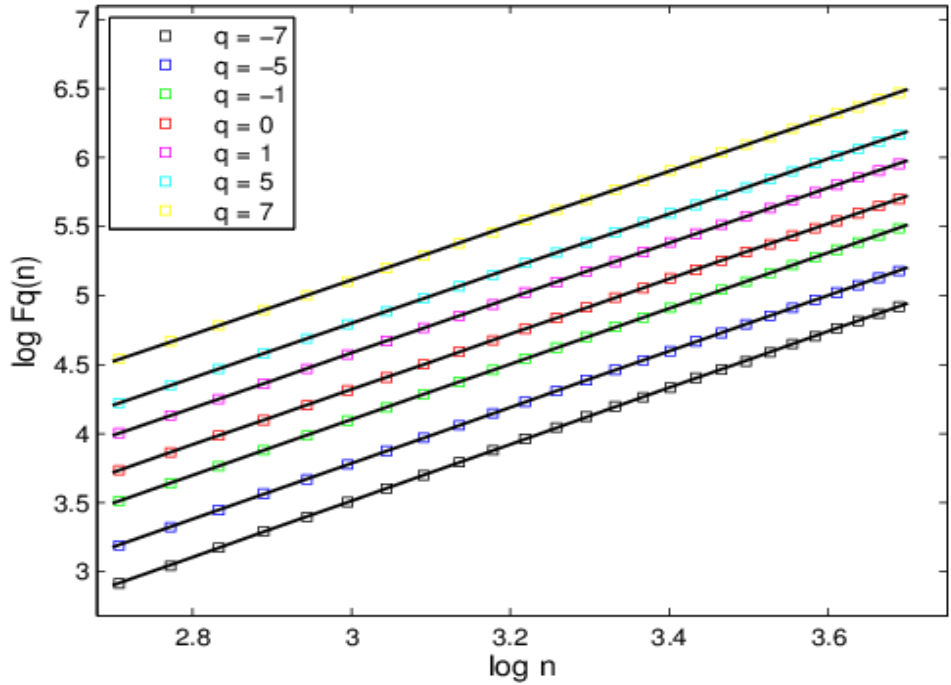

**Figure 2:** The plots of log Fq(n) versus log n for different values of q. The solid lines are the least squares fits to the data.

Other surfaces also exhibit excellent power law scaling (not shown here). The slopes obtained from the linear portions of $\log F_q(n)$ versus $\log n$ graphs for different values of $q$ are known as generalized Hurst exponents $h(q)$. In general, $q < 0$ corresponds to small fluctuations of $F_q(n)$ while for $q > 0$ the plot characterizes the large fluctuations [9].

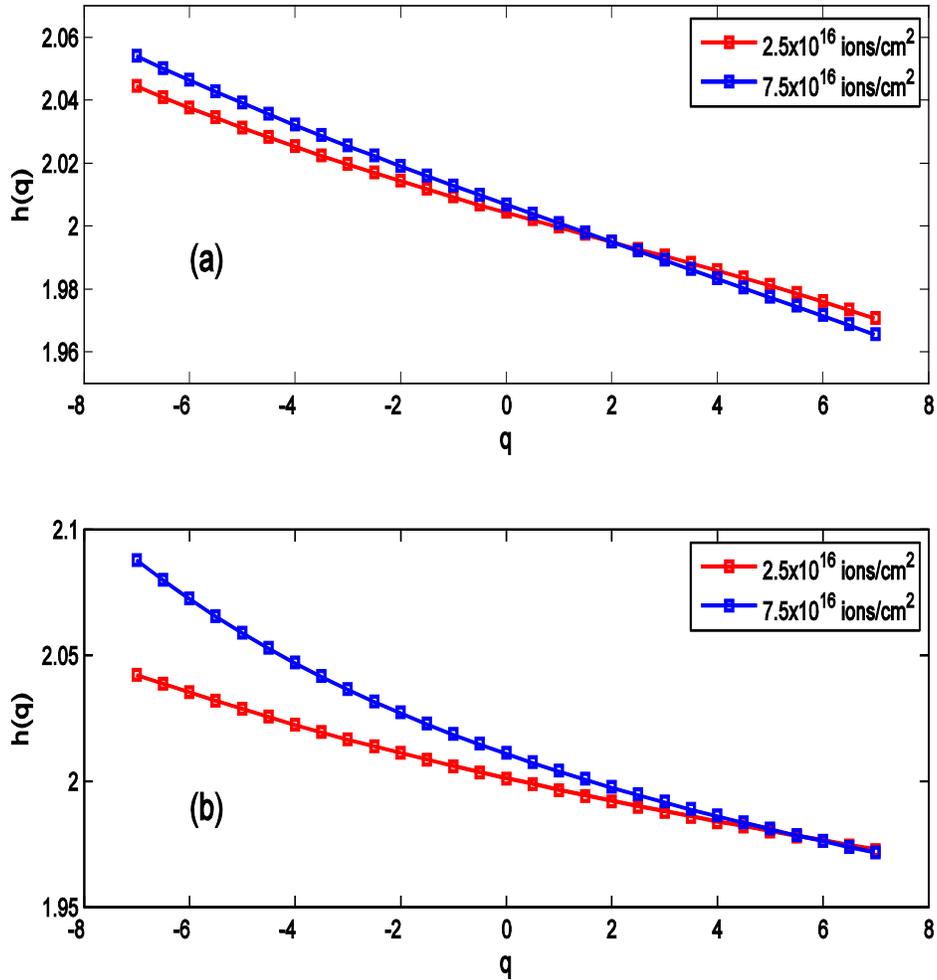

**Figure 3:** Scaling exponent $h(q)$ as function of $q$; before annealing (a) & after annealing (b)

The perfect fall of data points on straight lines are indicating the better choice of $q$ values and the surface under inspection have self-affine nature. For the larger value of $|q| > 7$ significant power law scaling is not observed or the scale invariant behavior is destroyed beyond $|q| > 7$. We plotted the graph between $h(q)$ and $q$ for each surface before and after annealing as shown in Figs. 3(a) and 3(b). Here, it is seen that $h(q)$ is a nonlinear decreasing function of $q$ for each surface. In Fig. 3(b), a greater nonlinearity is observed for the surface implanted with ions $7.5\times10^{16}$ ions/cm$^2$ after annealing. The nonlinear behavior of the surface suggested that the surface under investigation have multifractal nature. Therefore, nonlinearity is the hallmark of presence of multifractality. In addition, if $h(q)$ is $q$ dependent then surface is considered as multifractal; if not, it is monofractal. For positive $q$ values, $h(q)$ yields more stable results. It is found that the multifractality is enhanced after annealing.

The singularity strength, $\alpha(q)$ and the singularity spectrum, $f(\alpha)$ is computed by using the relation: $\alpha(q) = h(q) + qh'(q)$ and $f(\alpha) = q[\alpha - h(q)] + 2$ corresponding to each $q$ value [8]. These parameters are very important for characterization of multifractal surfaces. The measured values of these before and after annealing are listed in Table I and II. The multifractal spectra, the graphs between $f(\alpha)$ versus $\alpha(q)$ are shown in Figs. 4(a) and 4(b). Before annealing the shape of multifractal spectrum is shown as bell shaped, indicates that the height fluctuations of surface is closed [Fig. 4(a)] while transfer to left hooked after annealing, depicts that height fluctuations are varied [Fig. 4(b)]. The width of the multifractal spectrum ($\Delta\alpha$) is defined by $\Delta\alpha = \alpha_{max} - \alpha_{min}$ while the difference of singularity spectrum/fractal dimensions ($\Delta f$) are given by $\Delta f = f(\alpha_{min}) - f(\alpha_{max})$. A rapid increase for both parameters is noted for surface implanted with ions $7.5\times10^{16}$ ions/cm$^2$ after annealing. The larger value of $\Delta\alpha$ suggested the greater multifractality, increased roughness, coarser surface distribution, and increased non-uniformity of the height probabilities of surface. Here, $f(\alpha_{max})$ represents the fractal dimension corresponding to minimum growth probability while $f(\alpha_{min})$ represents maximum growth probability.

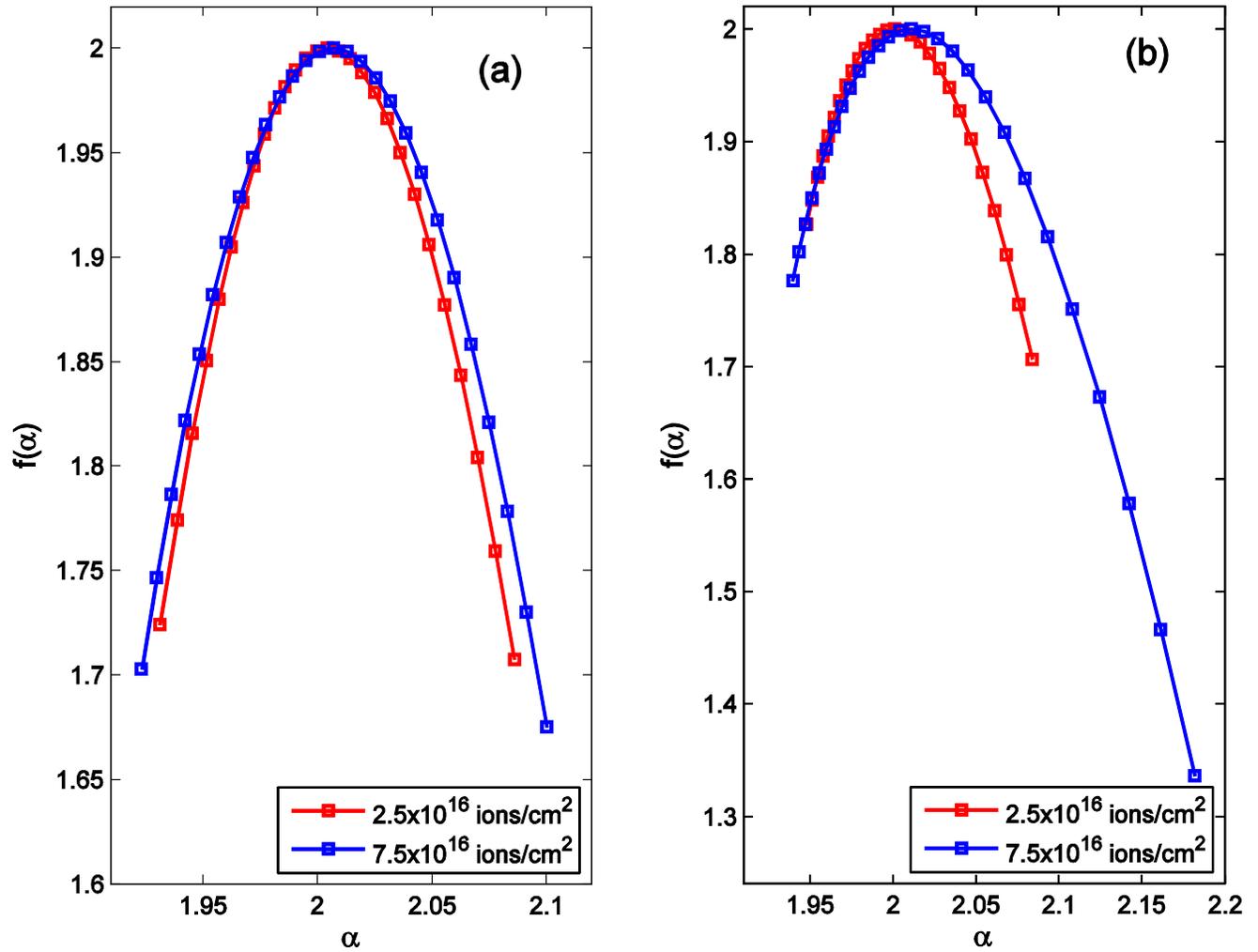

**Figure 4:** The multifractal spectra $f(\alpha)$ of surfaces; before annealing (a) & after annealing (b)

## 4. Conclusion:

AFM is used to capture the images of a SiO$_2$ film on a Si substrate implanted with 150 keV Ge ions at a fluences of $2.5\times10^{16}$ and $7.5\times10^{16}$ ions/cm$^2$ before and after annealing. Two dimension (2D) multifractal detrended fluctuation analysis (MFDFA) is applied to investigate the multifractality of each surface and found that each surface have multifractal nature. A nonlinear variation of generalized Hurst exponent with moment is the hallmark of presence of multifractality. Before annealing the shape of multifractal spectrum is shown as bell shaped, indicates that the height fluctuations of surface is closed, while transfer to left hooked after annealing, depicts that height fluctuations are varied. A rapid increase for both parameters ($\Delta\alpha$ and $\Delta f$) is noted for surface implanted with ions $7.5\times10^{16}$ ions/cm$^2$ after annealing. The larger value of $\Delta\alpha$ suggested the greater multifractality, increased roughness, coarser surface distribution, and increased non-uniformity of the height probabilities of surface.

## Acknowledgment:

RPY is thankful to Science and Engineering Research Board (SERB) of India for awarding National Postdoctoral Fellowship (PDF/2015/000590). VB is thankful to DST for providing funds under young scientist scheme (SR/FTP/PS-53/2011) and Nanomission scheme (IR/S2/PF/0001/2009).